\documentstyle[sprocl,epsfig,psfig]{article}
\def\beq{\begin{equation}}
\def\ra{\rightarrow}
\def\eeq{\end{equation}}

\def\bea{\begin{eqnarray}}
\def\eea{\end{eqnarray}}

\def\lam{\lambda}

\def\neut{{\tilde\chi}^0_1}

\newcommand{\gsim}{\stackrel{>}{\sim}}

\begin{document}

\vspace*{-35mm}
\begin{flushright}
hep-ph/9912217 \\
\hfill CERN-TH/99-375\\[-3ex]
\end{flushright}

\vspace{12mm}

\title{RESONANT SINGLE CHARGINO AND NEUTRALINO  \\
VERSUS FERMION-ANTIFERMION PRODUCTION \\ 
AT THE LINEAR COLLIDER
\footnote{Presented at the
$2^{nd}$ ECFA/DESY study on Linear Colliders, Frascati, 
    November 1998 (Alternative Theories working group).}}

\author{S. LOLA} 

\address{
CERN Theory Division, CH-1211, Geneva 23, Switzerland
}

\maketitle\abstracts{
We study single superparticle  productions
at the linear collider, putting particular
emphasis on resonant processes.
We find that there exists a wide region
of model parameters where single chargino and neutralino
productions dominate
over R-violating fermion-antifermion final states.
For certain values of $\mu$ and $M_2$,
it is possible to produce even the heavier charginos
and neutralinos at significant rates,
amplifying the total cross section
and obtaining interesting chains of cascade decays.
Effects from initial-state radiation are also included.
}


Although the main working tool for supersymmetry searches has 
been the Minimal Supersymmetric Standard Model,
the most general
$SU(3)_c\times SU(2)_L \times U(1)_Y$ invariant superpotential with the minimal
field content also contains the terms
\beq
W=\lam_{ijk} L_iL_j{\bar E}_k+\lam'_{ijk}L_iQ_j{\bar D_k}+
\lam''_{ijk}{\bar U_i}{\bar D_j}{\bar D_k}
\label{eq:superpot}
\eeq
where $L$ $(Q)$ are the left-handed lepton (quark) superfields while ${\bar
E},{\bar D},$ and ${\bar U}$ are the corresponding right-handed fields. 
If both lepton- and baryon-number violating operators  were present at
the same time in the low energy Lagrangian, they would lead to unacceptably
fast proton decay; to avoid this,
a symmetry that forbids the terms
in (\ref{eq:superpot}),
R-parity \cite{Fayet}, has been invoked. 
However, it has been shown that there exist
symmetries which allow the violation of only a subset of these operators
\cite{sym},
resulting in a very rich phenomenology \cite{pheno}:
single superparticle productions are allowed, while 
for couplings $\gsim10^{-6}$, the lightest supersymmetric 
particle decays inside the detector \cite{DRoss}.
In both cases, the standard missing energy signature is
substituted by multilepton and/or multijet events.

There are three basic categories of new signals: \\
$\bullet$
Pair superparticle productions and subsequent decays
via R-violating operators. Such processes
are favoured for small R-violating
couplings. \\
$\bullet$ For  reasonably large R-violating couplings,
single superparticle productions may
occur. In
 this case, 
the mass reach can be considerably larger than 
for MSSM processes at the same machine. \\
$\bullet$ Virtual effects, from sparticle exchanges.
These provide the optimal signals for a very
heavy superparticle spectrum.

Here, we put particular emphasis on
resonant scalar-neutrino production  
and its subsequent decay to either sfermions or
a single chargino or neutralino \cite{HH,herm,Kal,Feng}.
In particular, we study the processes
\begin{eqnarray}
e^{+} e^{-}  
 \rightarrow (\tilde{\nu})^* 
\rightarrow f \bar{f}' ~~~ {\rm and} ~~~~
e^{+} e^{-} \rightarrow (\tilde{\nu})^* 
 \rightarrow 
\left \{
\begin{array}{ll}
\ell^{\pm}_{i} \, 
\tilde{\chi}^{\mp}  \\
\nu_{i} \, \tilde{\chi}^{0} 
\end{array}
\right.
\nonumber
\label{eq:singleprod}
\end{eqnarray}
and identify for which regions of the
supersymmetric parameter space each 
channel is expected to dominate.

For a collider operating in the $e^+ e^-$ mode, 
the only couplings that involve two electrons are
$L_1L_2\bar{E}_1$ and $L_1L_3\bar{E}_1$
(remember that from $SU(2)$ invariance,
the two lepton doublets cannot have the same flavour)
\footnote{
In $2 \rightarrow 2$ single superparticle
productions one may generically probe only
a subset of operators. All 45 operators can be 
simultaneously probed by going to a 3-body final state,
for instance in rare $Z^{0}$ decays; 
however, this process is phase-space suppressed 
and was found to be more relevant for hadron colliders
\cite{MJ}.}.
The bounds for these couplings 
(Table 1) scale proportionally to the superparticle masses
\cite{constr,HH} and therefore,
for a heavy sparticle spectrum, the
couplings can be quite large.

\begin{table}
\begin{center}
\begin{tabular}{|lll|}
\hline
$ijk$ & ~~~~~$\lambda_{ijk}$ & ~~~~~~~~~Sources \\
\hline
 121 &  ~~~~~0.05 (0.5) &  ~~~~~~~~charged current universality \\
 131 &  ~~~~~0.06 (0.6) &  ~~~~~~~~$\Gamma(\tau
\rightarrow e\nu\bar{\nu})/ \Gamma(\tau 
\rightarrow \mu\nu\bar{\nu})$ \\
\hline
\end{tabular}
\caption{\normalsize \it
Upper limits on couplings for $m_{\tilde{f}} = 100 ~(1000) $ {\rm GeV}. }
\end{center}
\end{table}

Close to the resonance (where the $t$- and $u$- channel 
exchanges can be neglected in comparison to the $s$-channel pole), 
the cross section productions can be 
approximated by a Breit-Wigner formula. For instance,
for single neutralino production,
\begin{eqnarray}
        \sigma & = & \frac{8 \pi s}{m_{\tilde{\nu}}^2}
       \;
   \frac{\Gamma (\tilde{\nu} \rightarrow f \bar{f}) 
         \: 
         \Gamma (\tilde{\nu} \rightarrow \nu \tilde{\chi}_{0})
        }
        {   (s - m_{\tilde{\nu}}^2)^2 
          + m_{\tilde{\nu}}^2 \Gamma_{\rm total}^2 
        }
        \;
     \left[ \frac{ s - m_{\tilde{\chi}^0 }^2   }
                 { m_{\tilde{\nu}}^2 - m_{\tilde{\chi}^0 }^2   }
     \right]^2
        \nonumber\\
         & \rightarrow & \frac{8\pi}{m_{\tilde{\nu}}^{2}}
         B(\tilde{\nu}\rightarrow f \bar{f}) B(\tilde{\nu}\rightarrow 
         \nu\tilde{\chi}^{0})\;\hbox{, as $s\rightarrow m_{\tilde{\nu}}^{2}$}
        \label{BWneut}
\end{eqnarray}
Similar expressions arise for the other processes.
The resonant cross sections can thus be deduced 
by the appropriate branching 
fractions.  

Ignoring contributions to the vertices of the MSSM from mass terms, 
the latter are given by the following formulas:
\bea
\Gamma ({\tilde \nu} \to \nu \chi^0_i ) = 
\frac{g^2}{32 \pi}~ 
(N_{i2} - \tan \theta_{W} N_{i1})^2~ m_{\tilde{\nu}} \left
(1-\frac{m^2_{\chi^0_i}}{m_{\tilde{\nu}}^2} \right )^2 
\nonumber
\label{chirate}
\eea
\bea
\Gamma ({\tilde \nu} \to \ell^{\mp} \chi^\pm_i ) = 
\frac{g^2 V_{j1}^2}{16 \pi} 
 ~m_{\tilde{\nu}}~
(1-\frac{m^2_{\chi^\pm_i}}{m_{\tilde{\nu}}^2} )^2, 
~~~~\Gamma ({\tilde \nu} \to f \bar{f} ) = 
\frac{\lambda_{ijk}^2}{16 \pi} m_{\tilde{\nu}} \nonumber
\eea
In the above, $\lambda_{ijk}$ is the appropriate
R-parity violating Yukawa
coupling generating
the decay ${\tilde \nu} \to f \bar{f}$,while $V_{i1}$
and $N_{i1}$, $N_{i2}$ 
are the relevant matrix elements in the mixing matrix for
charginos and neutralinos respectively. 

Conclusively, whether single chargino and neutralino
final states will dominate over the
resonant fermion-antifermion productions depends on (i) the
SUSY parameter space and  (ii) the strength of $\lambda$.
This is indicated in Figs. 1,2 
where the branching ratio of the sneutrino decay 
to fermions is presented 
for the regions of the supersymmetry
parameter space that are interesting for
LEP (Fig.~1) and LC (Fig.~2) \footnote
{For squark decays, analogous results have been
presented in \cite{ContourPlots}.}.
Here, the U(1) gaugino mass $M_1$ is
determined from the SU(2) gaugino mass $M_2$
by the unification relation $M_1=(5/3)\tan^2\theta_WM_2$.
For lower values of $M_2,\mu$, 
a larger number of  charginos and
neutralinos
can be produced at the final state, while the
phase space suppression for their production 
is small. The picture starts changing as we pass to larger
$M_2,\mu$ and this is indicated in the
increase of the sneutrino decay rate to
fermions.  
However, we can see that for a wide range of
$M_2$ and $\mu$ the production of charginos and
neutralinos at the LC tends to dominate. 
Moreover, there exist bands of the
parameter space where the production of
the heavier charginos and neutralinos
may occur at a significant level.
This is shown  in Table 2, where we present the branching
ratios for the production of each chargino and neutralino
separately.

\begin{table}[h]
\centering
\begin{tabular}{|c|c|c|c|c|c|c|c|} \hline
{ $M_2$ } & 
{ $\mu$ } & 
{ ${ \Gamma_1}$ } &
{ ${ \Gamma_2}$ } &
{ ${ \Gamma_3}$ } &
{ ${ \Gamma_4}$ } &
{ ${ \Gamma'_1}$ } &
{ ${ \Gamma'_2}$ } \\
\hline \hline
 200. & -1000. &  0.53 &  1.49 &  -- &  -- &  2.93 &  -- \\
 200. &  -600. &  0.51 &  1.48 &  -- &  -- &  2.92 &  --\\
 200. &  -200. &  0.44 &  0.51 &  0.08 &  0.89 &  2.11 &  0.81\\
 200. &  200. &  1.09 &  0.28 &  0.01 &  0.58 &  2.09 &  0.88\\
 200. &  600. &  0.72 &  1.38 &  -- &  -- &  3.06 &  --\\
 200. &  1000. &  0.66 &  1.45 &  -- &  -- &  3.04 &  --\\
 \hline
 300. & -1000. &  0.49 &  0.85 &  -- &  -- &  1.68 &  -- \\
 300. &  -600. &  0.48 &  0.84 &  -- &  -- &  1.68 &  -- \\
 300. &  -200. &  0.46 &  0.01 &  0.06 &  0.75 &  0.41 &  1.30 \\
 300. &  200. &  0.82 &  0.01 &  0.01 &  0.50 &  0.92 &  0.87 \\
 300. &  600. &  0.62 &  0.85 &  -- &  -- &  1.88 &  -- \\
 300. &  1000. &  0.57 &  0.87 &  -- &  -- &  1.82 &  -- \\
 \hline
 400. & -1000. &  0.43 &  0.26 &  -- &  -- &  0.52 &  -- \\
 400. &  -600. &  0.42 &  0.26 &  -- &  -- &  0.52 &  -- \\
 400. &  -200. &  0.24 &  0.04 &  0.15 &  0.22 &  0.18 &  0.40 \\
 400. &  200. &  0.53 &  0.01 &  0.05 &  0.15 &  0.42 &  0.27 \\
 400. &  600. &  0.51 &  0.33 &  -- &  -- &  0.74 &  -- \\
 400. &  1000. &  0.48 &  0.30 &  -- &  -- &  0.63 &  -- \\
\hline
 500. & -1000. &  0.34 &  -- &  -- &  -- &  -- &  -- \\
 500. &  -600. &  0.34 &  -- &  -- &  -- &  -- &  -- \\
 500. &  -200. &  0.05 &  0.03 &  0.27 &  -- &  0.10 &  -- \\
 500. &  200. &  0.31 &  -- &  0.14 &  -- &  0.23 &  -- \\
 500. &  600. &  0.41 &  0.03 &  -- &  -- &  0.08 &  -- \\
 500. &  1000. &  0.38 &  -- &  -- &  -- &  0.01 &  --
\\ \hline
\end{tabular}

\caption{\normalsize \it 
All units in the table are in {\rm GeV}.
We chose
$tan\beta = 2$, $\lambda = 0.1$ and
$m_{\tilde{\nu}} = 500$ {\rm GeV}. 
$\Gamma_i$ are the decay rates for the
four neutralinos, while 
$\Gamma'_i$ the decay rates for the
two charginos.
For this choice of parameters,
the R-violating decay rate is
0.1 {\rm GeV}.}

\end{table}

\begin{figure}
\begin{minipage}[b]{8in}
\epsfig{file=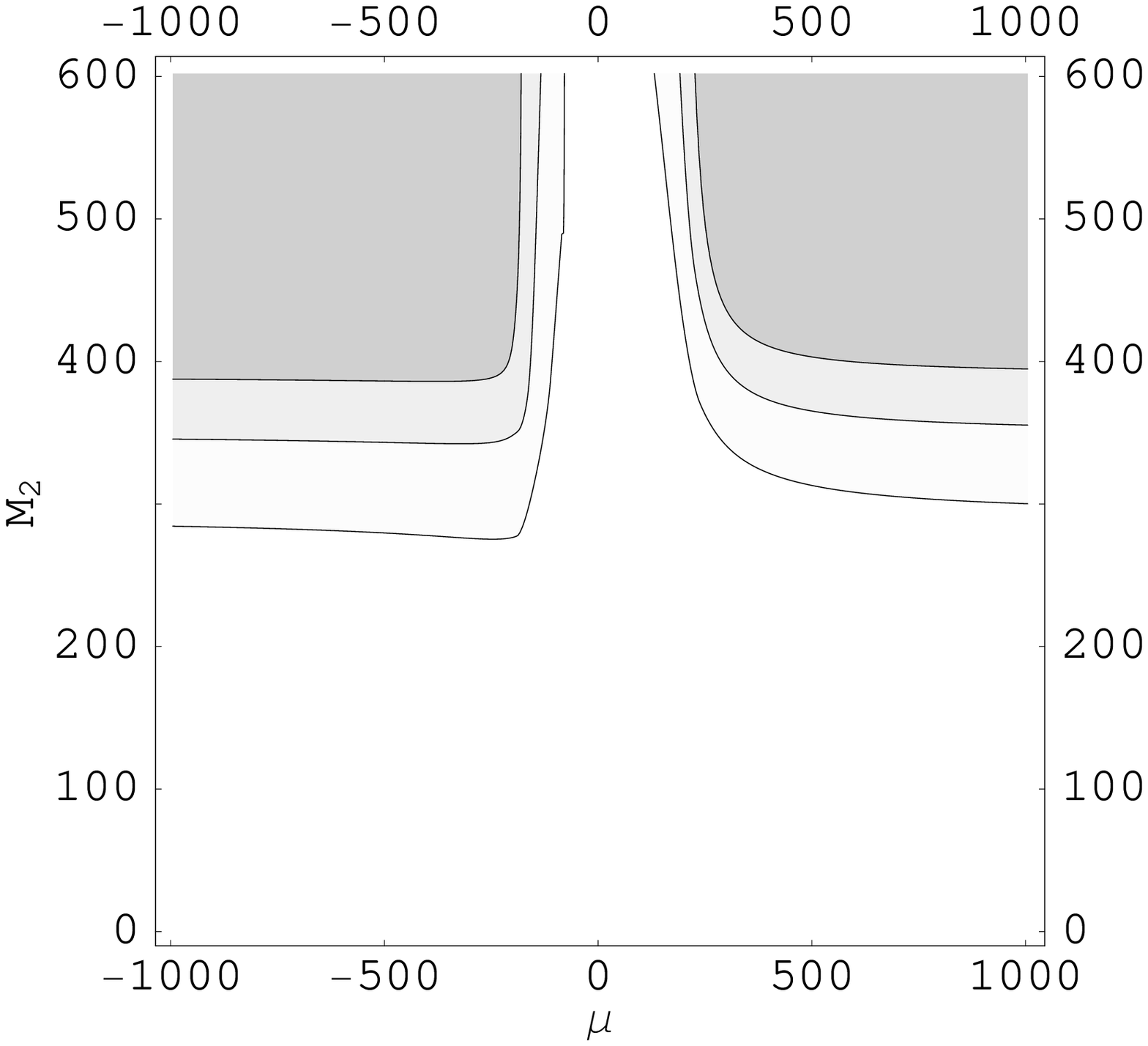,width=3in}
\epsfig{file=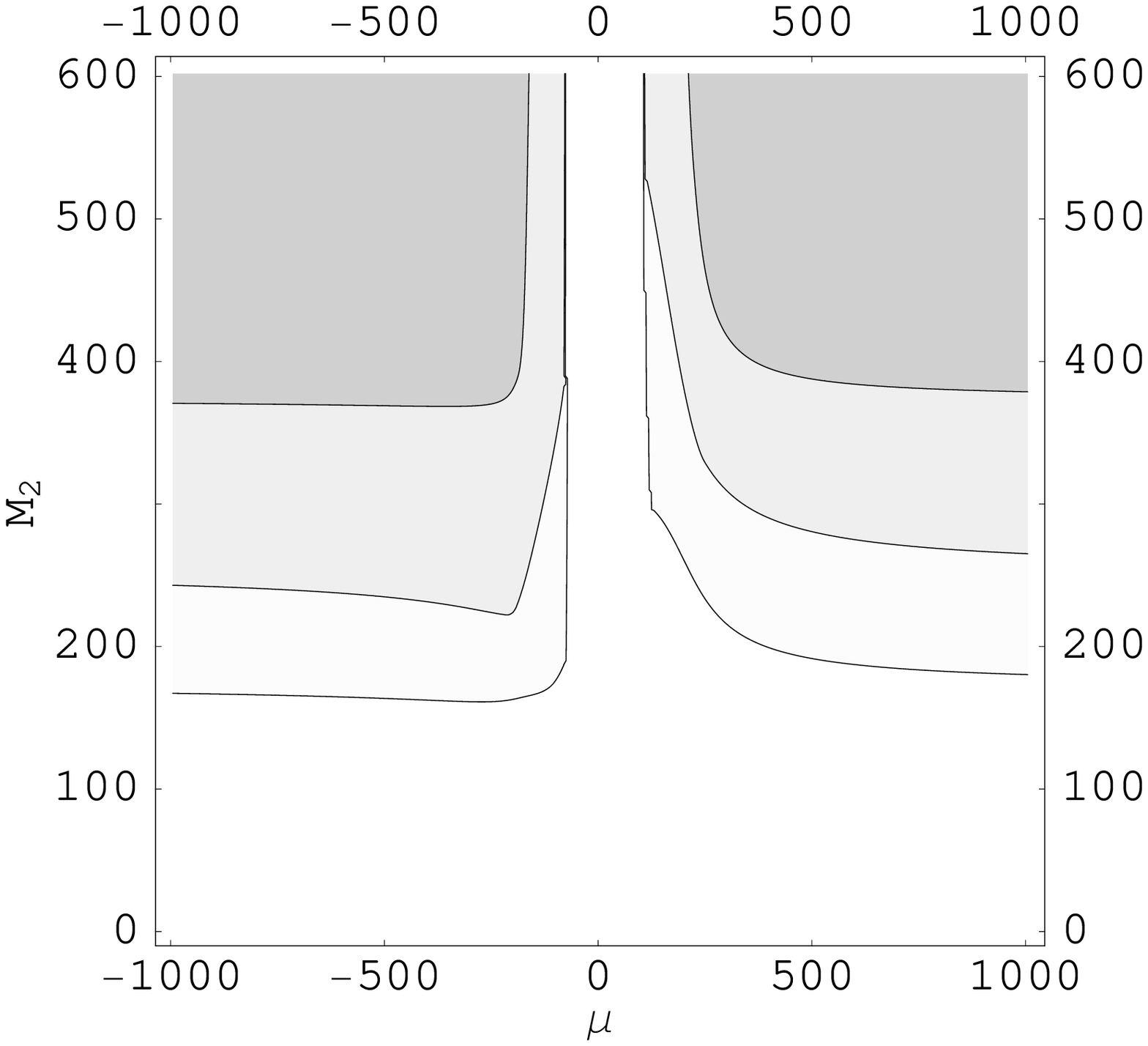,width=3in} \hfill
\end{minipage}
\vspace*{-3 cm}
\caption{\normalsize \it
${\cal B}$ for $\tilde{\nu}$ decay to fermions
for parameters most relevant for LEP.
We present contours for ${\cal B} \geq  0.9, 0.3, 0.1$,
from the darker to the lighter areas respectively.
We chose
$tan\beta = 2.0, m_{\tilde{\nu}} = 200$ {\rm GeV}
and $\lambda = 0.04$ (left) and 0.1 (right).
The LEP~2 bound on charginos
has been implemented.
}

\vspace*{0.5 cm}

\begin{minipage}[b]{8in}
\epsfig{file=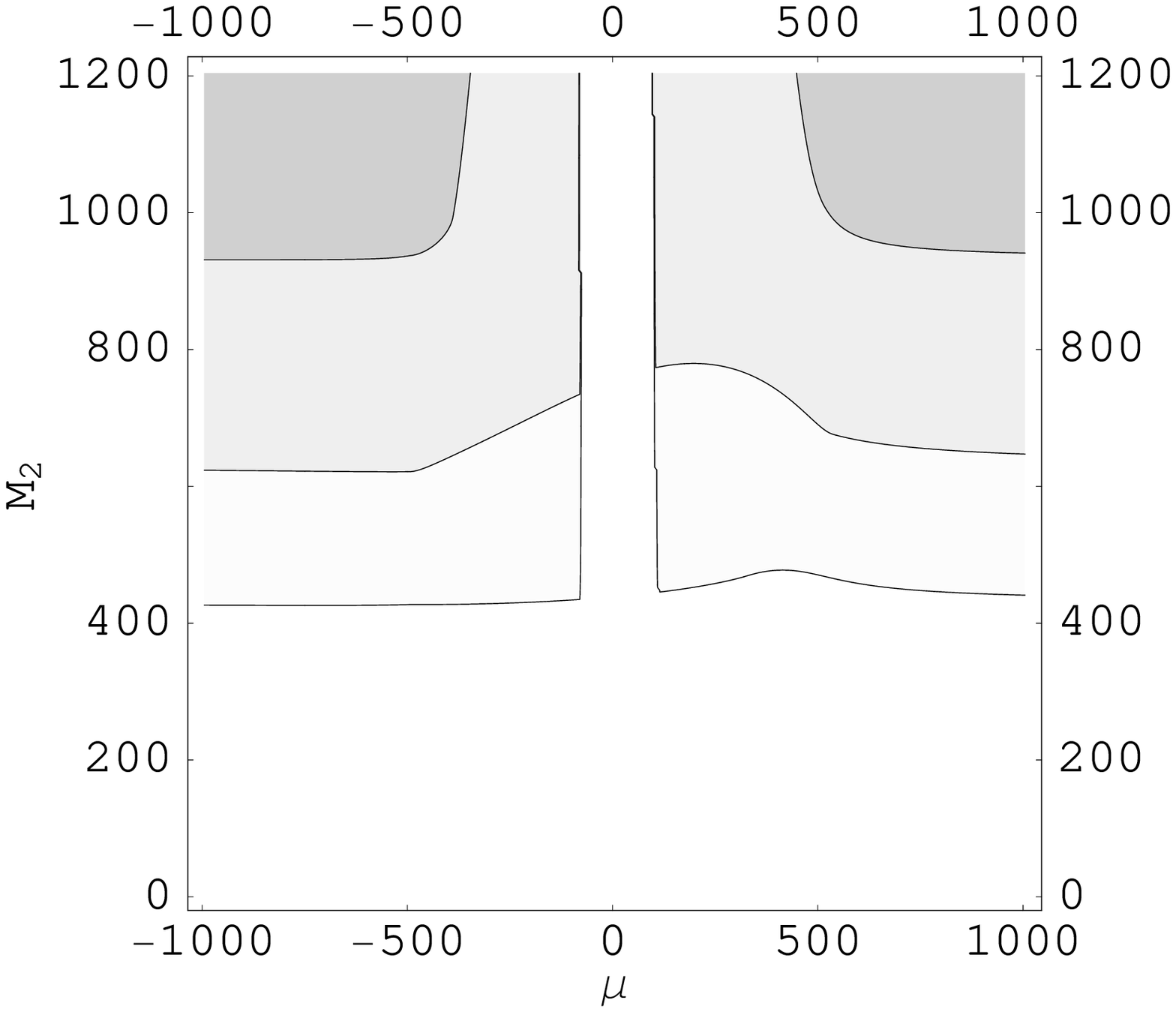,width=3in}
\epsfig{file=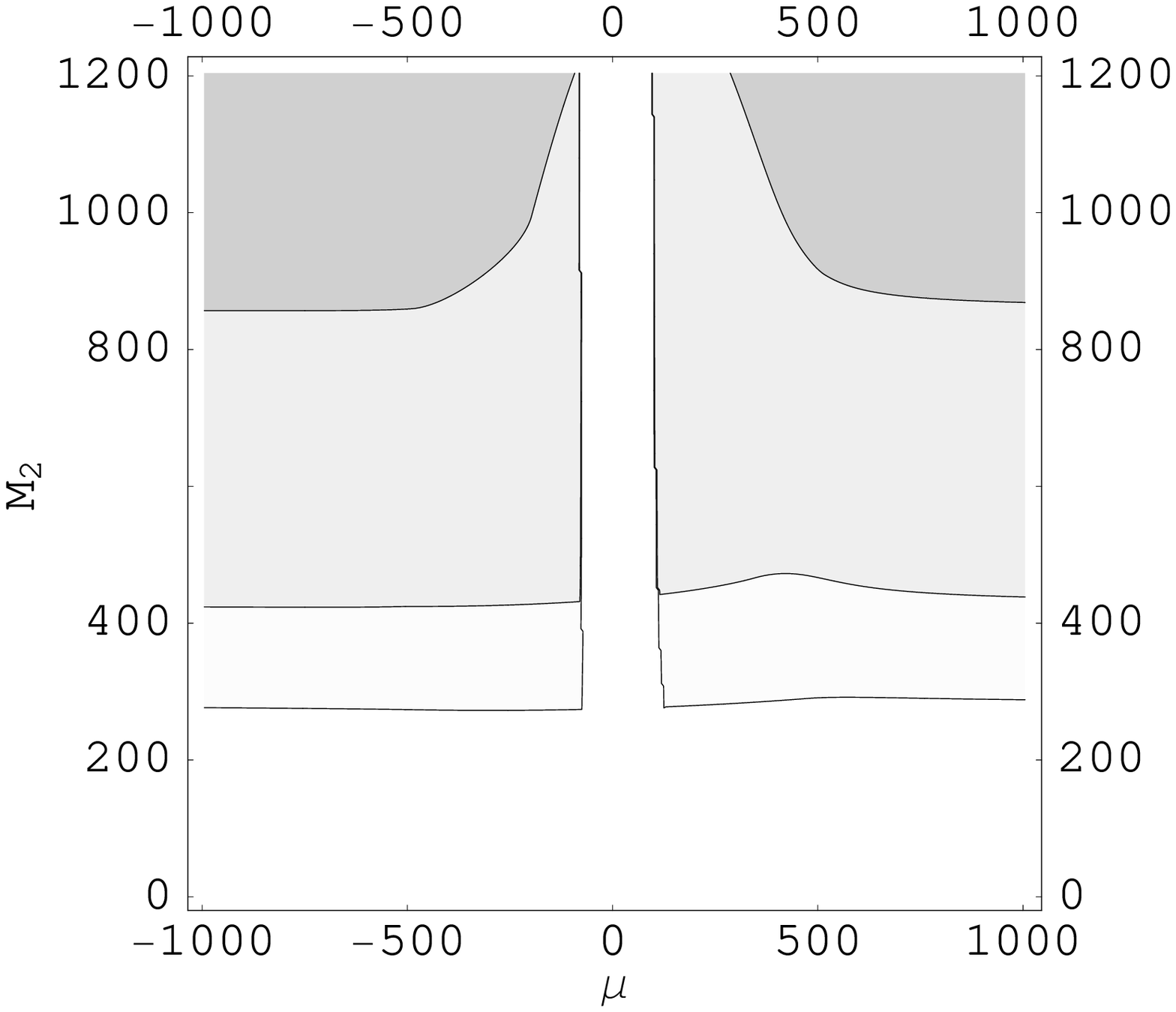,width=3in} \hfill
\end{minipage}
\vspace*{-3 cm}
\caption{\normalsize \it
${\cal B}$ for $\tilde{\nu}$ decay to fermions
for parameters most relevant for LC. We present
contours for ${\cal B} \geq 0.9, 0.3, 0.1 $,
from the darker to the lighter areas respectively.
We choose
$tan\beta = 2.0, m_{\tilde{\nu}} = 500$ 
{\rm GeV}
and $\lambda = 0.1$ (left) and 0.2 (right).
The LEP~2 bound on charginos
has been implemented.
}
\end{figure}

Subsequently, the charginos and neutralinos 
will decay to an R-parity even final state, with 
the possibility
of an interesting chain of 
cascade decays with multi-lepton
events and explicit
lepton-number violation at the final state.
The lightest neutralino decays via
\bea
\neut&\ra& \{(e^\pm,l_i^\mp,\nu_e),(e^\pm,e^\mp,\nu_i)\} 
\nonumber
\eea
For the charginos and the heavier neutralinos,
there exist two possible decay modes:
The first is the cascade decay via the lightest neutralino
and the second the direct decay via
the R-violating coupling(s), as discussed in \cite{hmp}.
For instance, for the lighter chargino we have
the channels
\bea
{\tilde\chi}^-_1\ra {\tilde \chi}_1^0+(W^-)^*\ra
{\tilde \chi}_1^0+f{\bar f}' \nonumber
\eea
where $f{\bar f}'$ are the decay fermions of the (virtual) W-boson,
or \beq
{\tilde\chi}^-_1\ra e^-e^+l_i^-,~~~~~~~~
{\tilde\chi}^-_1\ra \nu_e\nu_ie^- 
\label{dec}
\eeq
In the first case of (\ref{dec})
the total signal could be even more distinct since
it involves four leptons at the final state 
(three being in the
same semi-plane) without any missing energy, unlike the cascade
chargino decay which always involves neutrinos at the final state
\footnote{
Which of the two processes will appear, clearly depends on
(i) the strength of the R-parity violating operator:
the strongest the operator the larger the decay rate for 
a direct decay of the chargino.
(ii) the relative mass of chargino-neutralino:
if the mass gap between the two states is very small, then
the cascade decay is suppressed by phase space.
}. It turns out however that the charginos
as well as the heavier neutralinos dominantly
decay to $\chi^0_1$ and fermions for a wide
region of the parameter space.

In all cases, the signals should be clearly visible at an $e^+e^-$ collider,
provided the cross-section is sufficiently large;
the latter mainly depends on how large the unknown coupling $\lambda$ will be.
We study the relevant cross section, at and away 
from the resonance.
Ignoring contributions to the vertices of the MSSM from mass terms, 
we have two
channels present ($s$ and $t$) for chargino production
and all three  
($s$, $t$ and $u$) for neutralino production.
For the $s$-channel
diagram we take into account the contribution due to the 
decay width of the scalar neutrino.

\begin{figure}[tbp]
\begin{center}
\epsfig{file=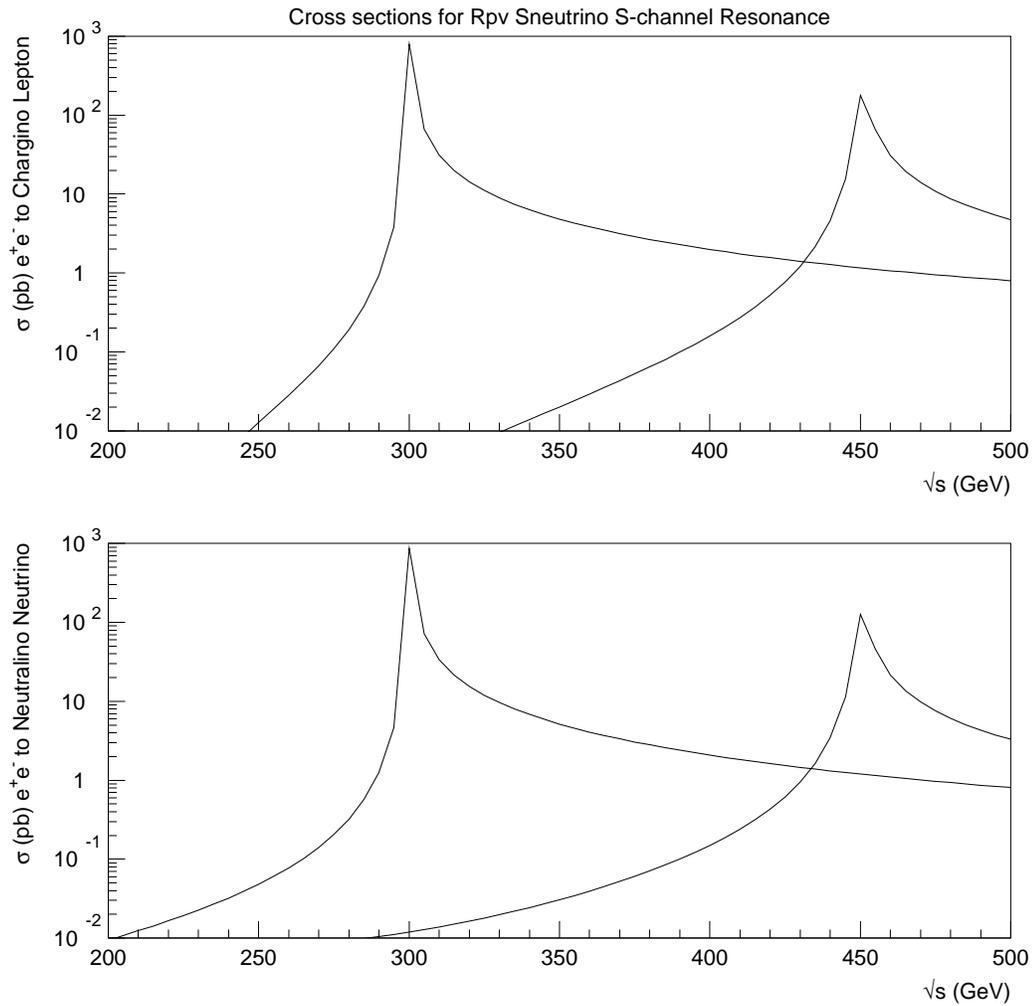,width=15cm}
\end{center}
\caption{\normalsize \it
The parameters for this plot are the following:
$m_{\tilde{\nu}} = 300$ and 450 {\rm GeV} respectively;
$m_{\tilde{e}} = 1000$ {\rm GeV};
$\tan\beta = 2, \lambda = 0.1,
\mu = -200 ~{\rm GeV}, M_2 = 250$
{\rm GeV}. In this case, all
charginos and neutralinos are produced.
}
\end{figure}

In Fig.~3, we show the cross sections for single
chargino and neutralino productions, including effects
from initial state radiation (ISR),
for two different sneutrino masses.
To illustrate the effects
from the production of many charginos and neutralinos we
chose a point of the parameter space where
all several of these states 
are produced.
Indeed, for the choice of parameters
tha appears in the figure, the chargino masses are
201.9 and 273.1 GeV respectively,
while the neutralino masses are
127.8, 192.9, 217.5 and 272.1 respectively.

As expected,
the effect of initial state radiation is to lower the peak
but widen the resonance. For instance, in our example
we find that, for $\sqrt{s} = 500$ GeV
(where all charginos and neutralinos may be
produced)
and $m_{\tilde{\nu}} = 450$ GeV, IR enhances
the cross section by almost an order of magnitude.
Actually, in this example,
the heavier charginos and
neutralinos may arise with large cross sections.
Indeed, the partial cross sections that we find
for the four neutralinos
(from the lighter to the heavier), for
$\sqrt{s} = 500$ and $m_{\tilde{\nu}} = 450$ GeV, are:
1.03 pb, 0.22 pb, 0.15 pb, 1.9 pb
while for the charginos 1.8pb and 2.9 pb 
respectively.

From the above discussion,
we conclude that single chargino and neutralino
productions arise with significant cross sections
and provide
an interesting possibility for looking for
R-violating supersymmetry at the Linear Collider.

\vspace*{0.2 cm}
{\bf Acknowledgement:} I would like to thank P. Morawitz
for collaborating at the first stages of this work
and A. Belyaev, for his help with PAW.

\end{document}